\documentclass{article}

\usepackage{graphicx}
\usepackage{amsmath, amssymb}
\usepackage{enumerate}
\usepackage{times}

\begin{document}

\title{Number sequence representation of protein structures based on the second derivative of a folded tetrahedron sequence}

\author{Naoto Morikawa (nmorika@genocript.com)}

\date{October 7, 2006.}

\maketitle


\begin{abstract}
  A protein is a sequence of amino-acids of length typically less than $1,000$, where there are $20$ kinds of amino-acids. In nature, each protein is folded into a well-defined three-dimensional structure, the native structure, and its functional properties are largely determined by the structure. Since many important cellular functions are carried out by proteins, understanding the native structure of proteins is the key to understanding biology at the molecule level.
 
This paper proposes a new mathematical approach to characterize native protein structures based on the discrete differential geometry of tetrahedron tiles.  In the approach, local structure of proteins is classified into finite types according to shape. And one would obtain a number sequence representation of protein structures automatically.  As a result, it would become possible to quantify structural preference of amino-acids objectively. And one could use the wide variety of sequence alignment programs to study protein structures since the number sequence has no internal structure. 

The programs are available from \texttt{http://www.genocript.com}. 
\end{abstract}



\section{Introduction}

  A protein is a sequence of amino-acids of length typically less than $1,000$, where there are $20$ kinds of amino-acids. In nature, each protein is folded into a well-defined three-dimensional structure, the native structure, and its functional properties are largely determined by the structure. Since many important cellular functions are carried out by proteins, understanding the native structure of proteins is the key to understanding biology at the molecule level.

Currently protein structures are characterized by classifications based on structural similarity. 
But classification is to some extent subjective and there exists a number of classification databases with different organization, such as CATH \cite{CATH}  and SCOP \cite{SCOP}. For example, protein structures are usually described using intermediate structure, such as $\alpha$-helix, $\beta$-sheet, and turn, which are formed by hydrogen-bonding between distant amino-acids. But there exists no consensus about the assignment of secondary structure, particularly their exact boundaries. 
(In protein science the intermediate structure is referred to as secondary structure, whereas the amino-acid sequence is primary and the spatial organization of secondary structure elements is tertiary.)

This paper proposes a new mathematical approach to characterize native protein structures based on the discrete differential geometry of tetrahedron tiles \cite{M}.  In the approach, local structure of proteins is classified into finite types according to shape. And one would obtain a number sequence representation of protein structures automatically.  As a result, it would become possible to quantify structural preference of amino-acids objectively. And one could use the wide variety of sequence alignment programs to study protein structures since the number sequence has no internal structure.

The programs are available from \texttt{http://www.genocript.com}. 

\section{Previous works}
An amino-acid sequence has only two rotational freedom per each amino-acid and protein structures was often represented by the two rotational angles, referred as the Ramachandran plot. The torsion angle between C$\alpha$ atoms were also used to define second structure \cite{L}.

\cite{RS} proposed a representation of protein structures based on differential geometry. In their method, a protein is represented as broken lines and they defined the curvature and torsion at each point. And \cite{RF} described the topology of a protein by $30$ numbers inspired by knot theory.

Finally, \cite{RR} defined an imaginary cylinder to describe helices, whose axis is approximated by calculating the mean three-dimensional coordinate of a window of four consecutive C$\alpha$ atoms. And \cite{CCL} extended the result to define secondary structure based on a number of geometric parameters.

For more information, see \cite{TA} which reviews various geometric methods for non-(protein) specialists.

As for one-dimensional profile of protein structures, \cite{BLE} described every amino acid position in terms of its solvent accessibility, the polarity of its environment and its secondary structure location. And \cite{JTT} used energy potential to describe amino acid positions.

On the other hand, \cite{T} proposed a ``periodic table'' constructed from idealized stick-figures as a classification tool, which shifts the classification from a clustering problem to that of finding the best set of ideal structures. And \cite{TSS} proposed a set of short structural prototypes,  ``structural alphabet'', to encode protein structures.

\section{Encoding method}

\subsection{Differential structure of a tetrahedron sequence}

\begin{figure*}[tb]
\includegraphics[scale=0.9]{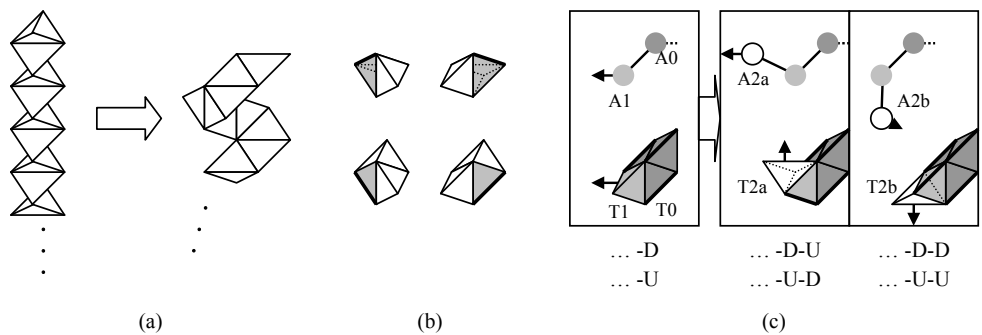}
\caption{Tetrahedron sequence.
(a): Folding.
(b): Four directions of a tile (gray). 
(c): Coding rule (Assignment of the ``second derivative'').}
\label{fig1}
\end{figure*}

We approximate native protein structures by a particular kind of tetrahedron sequence which satisfies the following conditions  (Fig.\ref{fig1}(a)): (i) each tetrahedron consists of four short edges and two long edges, where the ratio of the length is $\sqrt{3}/2$  and (ii) successive tetrahedrons are connected via a long edge and have the rotational freedom around the edge.

Each tetrahedron of a folded sequence assumes one of the four directions of its short edges, which is determined by the configuration of the previous and next tiles (Fig.\ref{fig1}(b)). Then, we can describe change of the direction of tiles, i.e. the ``second derivative'', by a binary sequence, where either $U$ or $D$ is assigned to each tile. The coding rule is simple: change the value if the direction changes.

For example, suppose that a protein backbone has been approximated up to the previous atom $A0$ by a tetrahedron sequence up to tetrahedron $T0$ (Fig.\ref{fig1}(c)). That is, the direction of $T0$ and the position of $T1$ has been determined by the position of the current atom $A1$ (left). Then, there are two candidates $T2a$ and $T2b$ for the position of the next tile $T2$. 
If the next atom is $A2a$, then $T2a$ is closer to the atom. Thus the next tile assumes the position of $T2a$ and the direction of tetrahedrons changes (middle). In this case assign $U$ to $T1$ if the value of $T0$ is $D$ and assign $D$ to $T1$ otherwise. On the other hand, if the next atom is $A2b$, then the next tile assumes the position of $T2b$ and the direction of tetrahedrons does not change (right). In this case assign $D$ to $T1$ if the value of $T0$ is $D$ and assign $U$ to $T1$ otherwise. At endpoints, choose the tile which is closer to $A1$. (See \cite{M} for the mathematical foundation.)

\subsection{$5$-tile code of proteins}

\begin{figure*}[tb]
\includegraphics[scale=0.9]{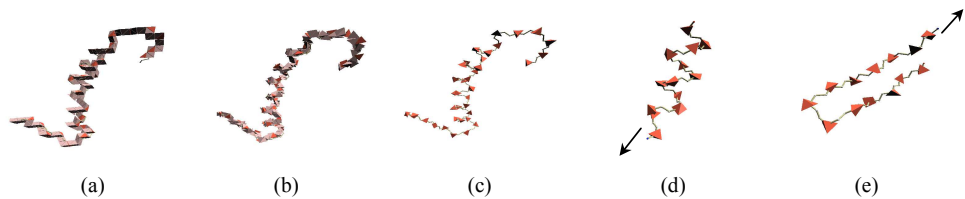}
\caption{Approximation of transferase, PDB ID $1$rkl, and others. Broken lines show protein backbones.
(a): Folding only. 
(b): Folding with rotation. 
(c): Folding with rotation and translation. 
(d): $\alpha$-helix (from the $142$-th to $154$-th amino-acids of protein $1$be$3$).
(e): $\beta$-sheet (from the $826$-th to $838$-th amino-acids of protein $1$jz$7$).}
\label{fig2}
\end{figure*}

Upon approximation, we consider the position of the centre of amino-acids of proteins. In other words, we identify an amino-acid with the $\alpha$-carbon atom (C$\alpha$) located in its centre. And we allow rotation and translation of tetrahedrons during the folding process to absorb irregularity of actual protein structures. That is, once the direction of tile $T1$ is determined  (Fig.\ref{fig1}(c)), we rotate $T1$ to make its direction parallel with the direction from $A0$ to $A2$ and translate $T1$ to the position of $A1$. 

Fig.\ref{fig2}(a), (b), and (c) show approximation of a protein (transferase), whose PDB (Protein Data Bank) ID is $1$rkl, by folding only, by folding with rotation, and by folding with rotation and translation respectively.

To study protein structures, we consider every fragment of an bodd number of amino-acids, say $n$, contained in a given protein. We start encoding from the middle point amino-acid, say $A$, and call the obtained binary sequence \textit{$n$-tile code of amino-acid $A$}, where $A$ is always assigned value $D$. (Note that encoding depends on choice of the initial amino-acid.) 
For example, the middle point amino-acid of the $\alpha$-helix shown in Fig.\ref{fig1}(d) is encoded into $13$-tile code $D$$D$$D$$D$$D$$U$$D$$U$$D$$D$$D$$D$$D$ and 
the middle point amino-acid of the $\beta$-sheet shown in Fig.\ref{fig1}(e) is encoded into 
$D$$D$$D$$D$$D$$D$$D$$D$$D$$D$$D$$D$$D$ (all Ds!).

Since it is enough to consider $5$-tile codes to detect $\alpha$-helix, we restrict ourselves on $5$-tile codes below.
Using $5$-tile codes, we obtain $16$-valued sequence for each protein. For example, the $\alpha$-helix of Fig.\ref{fig1}(d) is encoded into a $5$-tile code sequence of length nine, 
$H$$H$$H$$H$$H$$H$$H$$H$$H$ , and the $\beta$-sheet of Fig.\ref{fig1}(e) is encoded into 
$S$$S$$S$$T$$S$$S$$S$$S$$S$, where $H$ stands for $DUDUD$, $S$ for $DDDDD$, and $T$ for $UUDUU$. See appendix A for the list of $5$-tile codes and their examples.

\section{Results}

\subsection{$5$-tile code assignment of superfolds}

\begin{table*}[tb]
\includegraphics{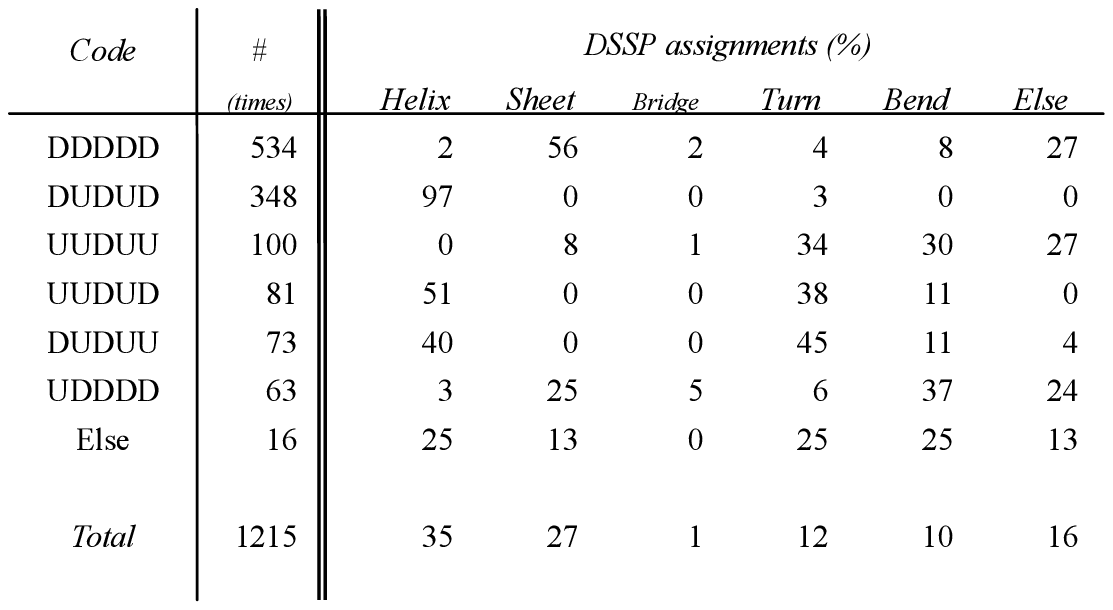}
\caption{Frequency of $5$-tile codes and their DSSP assignments (superfolds).}
\label{tab1}
\end{table*}

It is well known that there exist highly populated families of second structure arrangements, called superfolds \cite{OJT}, shared by a diverse range of amino-acid sequences. And we consider nine superfolds, from which we extract $1,215$ fragments of $5$ amino-acids, to study the correspondence between $5$-tile codes and secondary structure of proteins, such as helix, sheet, and turn.
We use DSSP (Dictionary of Secondary Structure of Proteins) program \cite{BT} to assign secondary structure elements to each amino-acid, which calculates energies of hydrogen bonds using a classical electrostatic function.

Table \ref{tab1}(a) shows the frequency distribution of $5$-tile codes and their DSSP assignments. This shows that code $DDDDD$ mainly corresponds to sheet, $DUDUD$ mainly to helix, 
and $UUDUU$ mainly to turn. On the other hand, most of helixes are covered by three codes $DUDUD$, $UUDUD$, and $DUDUU$, most of sheets are covered by $DDDDD$, and most of turns are covered by $UUDUU$, $UUDUD$, and $DUDUU$.

Appendix B shows the spacial distribution of four $5$-tile code groups, mainly-helix \{DUDUD\}, mainly-sheet \{DDDDD\}, mainly-turn \{UUDUU, UUDUD, DUDUU\}, and else.

\subsection{Frequency distribution of $5$-tile codes}

\begin{table*}[tb]
\includegraphics[scale=0.9]{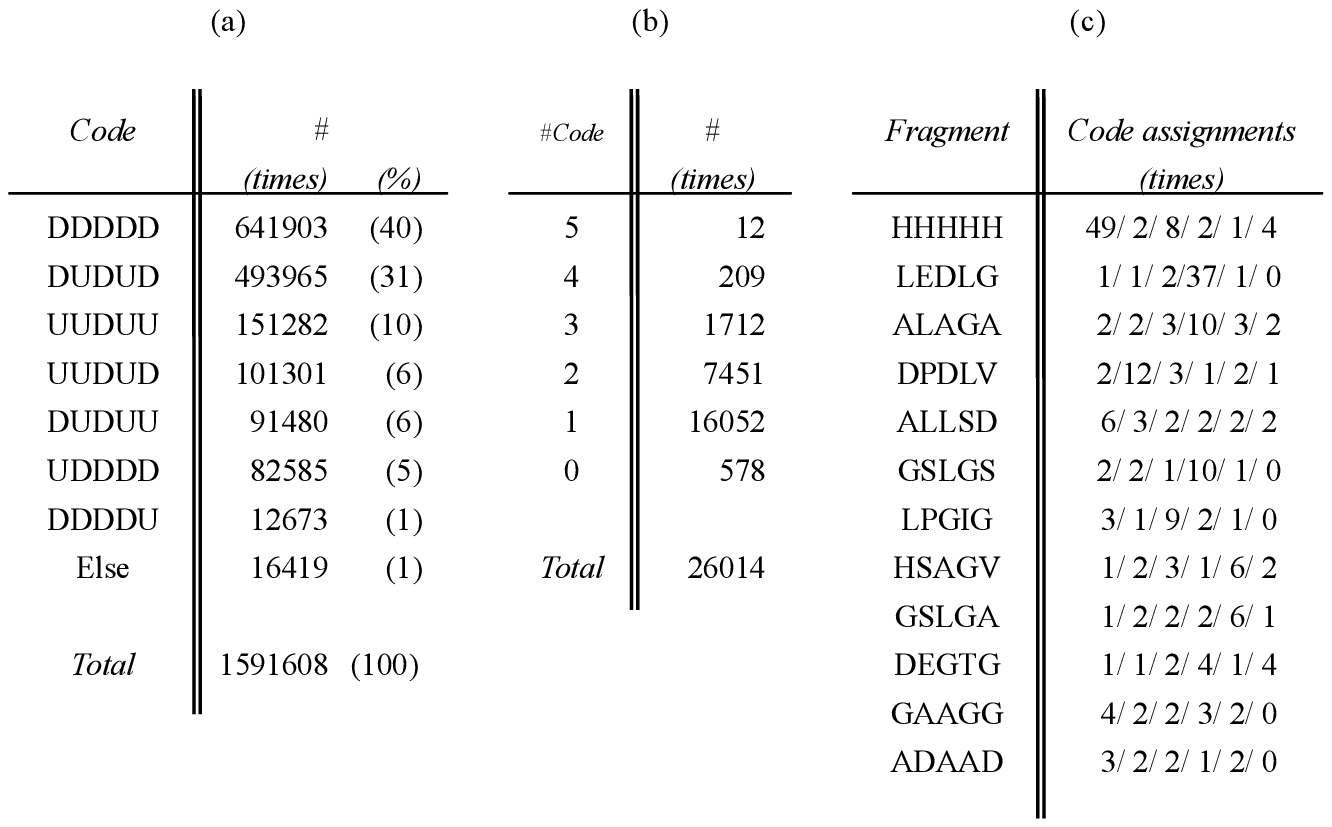}
\caption{Statistics of $5$-tile codes. 
(a): Frequency of $5$-tile codes. 
(b): Frequency of amino-acid sequences which relate to a given number of the top five $5$-tile codes. (Sequences which occurred more than nine times only are considered)
(c): $5$-tile code assignments of the amino-acid sequences which relate to all of the top five $5$-tile codes. Sequences are given by one-letter code of amino-acids and the figures show the frequency of $DDDDD$/$DUDUD$/$UUDUU$/$DUDUU$/$UUDUD$/Else.}
\label{tab2}
\end{table*}

Currently protein structures are classified in a hierarchical fashion. For example, the SCOP (structural classification of proteins) database classifies more than $25,000$ proteins into about $3,000$ families manually.

To study the frequency distribution of $5$-tile codes, we took one protein for each SCOP family ($1.69$ release), from which we extracted $1,591,608$ fragments of $5$ amino-acids. They corresponds to $612,232$ types of  amino-acid sequences, where $26,014$ types are occurred more than nine times.

As table \ref{tab2}(a) shows, $93$\% of the fragments are covered by top five $5$-tile codes. 
Table \ref{tab2}(b) is concerned with the number of the codes related to one amino-acid sequence, where sequences which occurred more than nine times only are considered. For example, more than $60$\% of the amino-acid sequences are encoded uniquely and there exist twelve sequences which relate to all of the five codes. Table \ref{tab2}(c) shows their $5$-tile code assignments. Note that most of them have a preference for a specific code.

Finally, appendix C gives the preference of amino-acids for $5$-tile codes. As you see, amino-acids are grouped into three categories: $DDDDD$-oriented, $DUDUD$-oriented, and others.

\section{Discussion}

Recently more than a few secondary structure assignment programs are available. But  assignments differ from one program to another because of the arbitrariness of the definition of secondary structure, particularly its exact boundaries. According to \cite{CCL}, the degree of disagreement between two different assignments could reach almost $20$\%.

On the other hand, our approach dose not require any pre-defined secondary structure. Instead, it classifies local structure according to shape. For example, in the case of $5$-tile coding, local structure is grouped into $16$ types of elements (See appendix A). 
Though it can neither distinguish three types of helixes from one another nor describe global features directly, such as hydrogen bonding, it can detect secondary structure to same extent (See appendix B). That is, it allows a comprehensive description of, not only specific (secondary) structure, but also arbitrary local structure. And it becomes possible to quantify structural preference of amino-acids objectively (See appendix C).  It could be used to characterize binding sites of drugs and the active site of enzymes, too.

Moreover, since our method encodes a protein into a number sequence without any structure, 
it is easy to analyse the (number sequence) profile and we could use the wide variety of sequence alignment programs to compare protein structures. And the incorporation of structural information should lead to more powerful protein structure predictions because structure is much more strongly conserved than sequence during evolution. For example, in the case of $5$-tile coding, more than $60$\% of amino-acid sequences are uniquely encoded. Thus, they can be substituted with $5$-tile codes when one considers plausible structures of a given amino-acid sequence.

\section{Conclusions}

This paper proposes a new mathematical approach to characterize native protein structures based on the discrete differential geometry of tetrahedron tiles \cite{M}.  In the approach, local structure of proteins is classified into finite types according to shape. And we have obtained a comprehensive description of, not only specific (secondary) structure, but also arbitrary local structure elements. As a result, one could quantify structural preference of amino-acids objectively. And one could use the wide variety of sequence alignment programs to study protein structures since the one-dimensional profile of the assignment is given as a simple number sequence.



\clearpage

\appendix

\renewcommand{\thesection}{\Alph{section}}
\section{$5$-tile code list}\label{app1}

\begin{table*}[hb]
\includegraphics[scale=0.7]{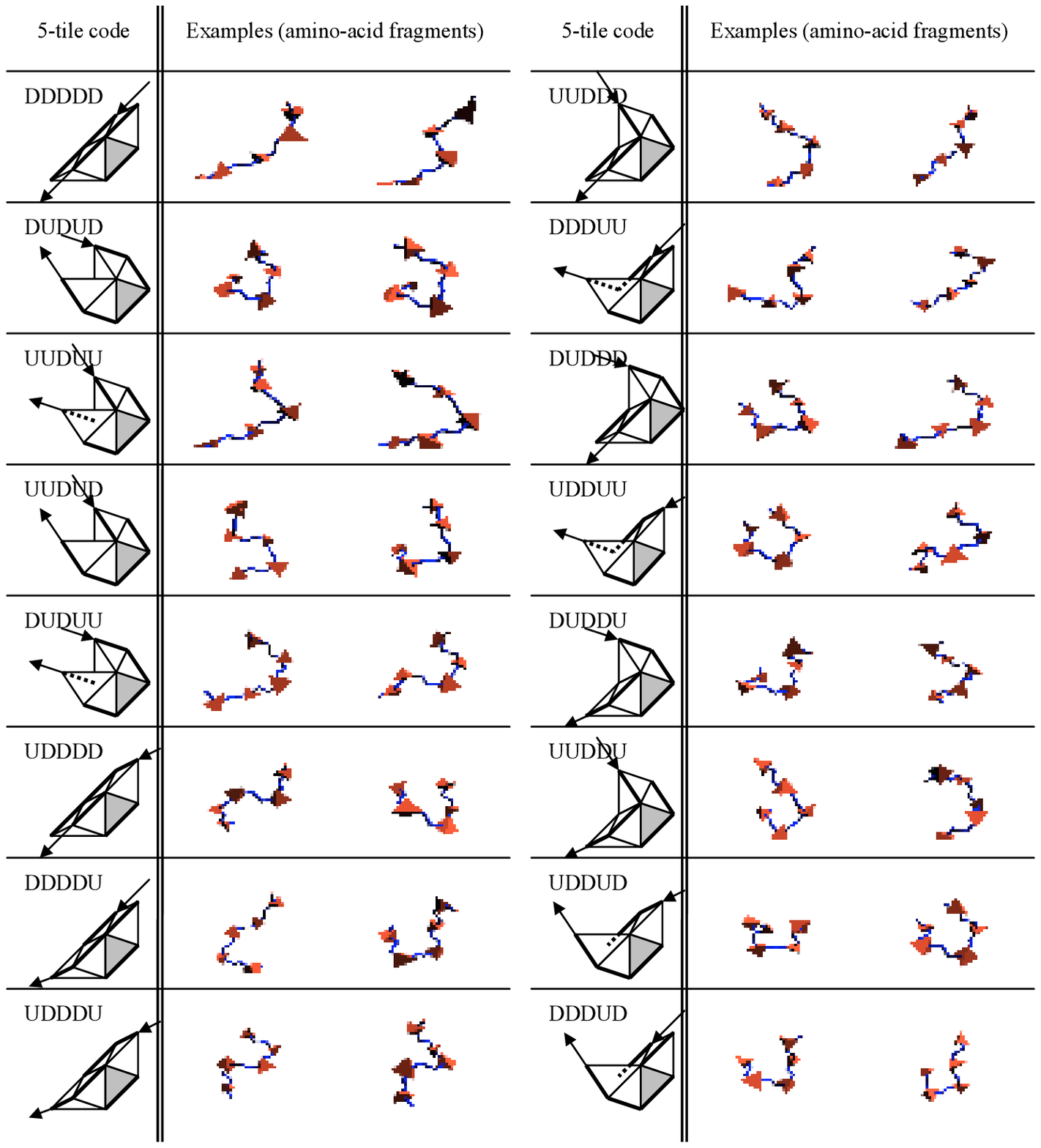}
\caption{$5$-tile code list. Broken lines show protein backbones.}
\label{tab3}
\end{table*}

\clearpage

\section{Spacial distributions of $5$-tile codes}

\begin{figure*}[hb]
\includegraphics[scale=0.8]{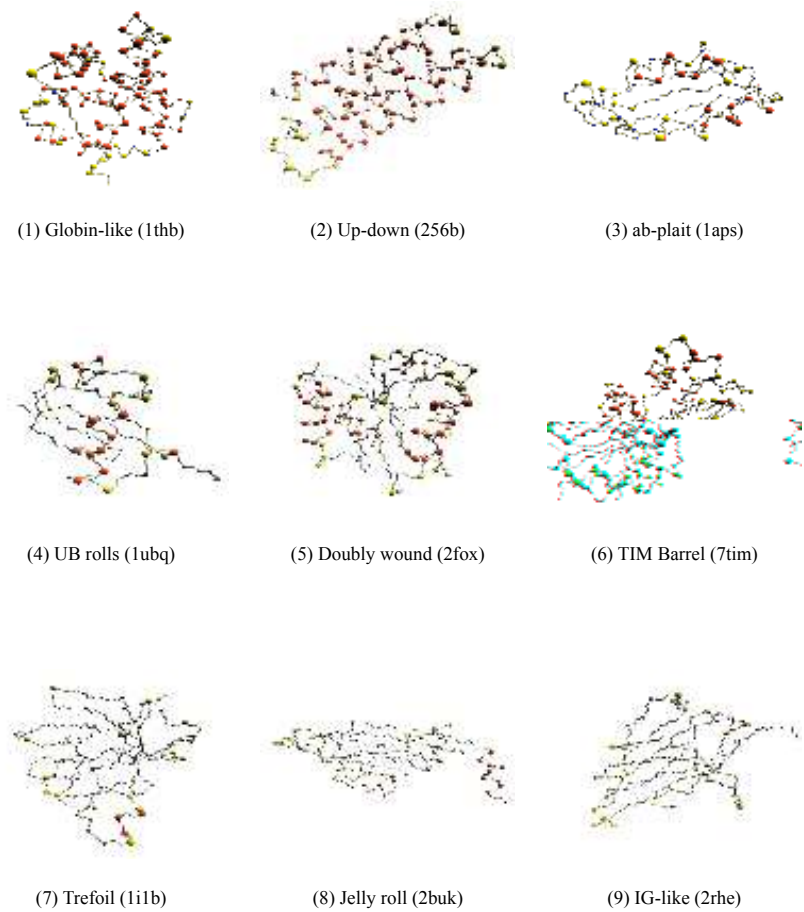}
\caption{$5$-tile code distributions of nine superfolds. Broken lines show protein backbones and 
Red balls are the C$\alpha$ atoms of the amino-acids of mainly-helix code $DUDUD$, yellow balls mainly-turn codes \{$UUDUU$, $DUDUU$, $UUDUD$\}, and small blue balls mainly-sheet code $DDDDD$.}
\label{fig3}
\end{figure*}

\clearpage

\section{$5$-tile code preference of amino-acids}

\begin{table*}[hb]
\includegraphics[scale=0.9]{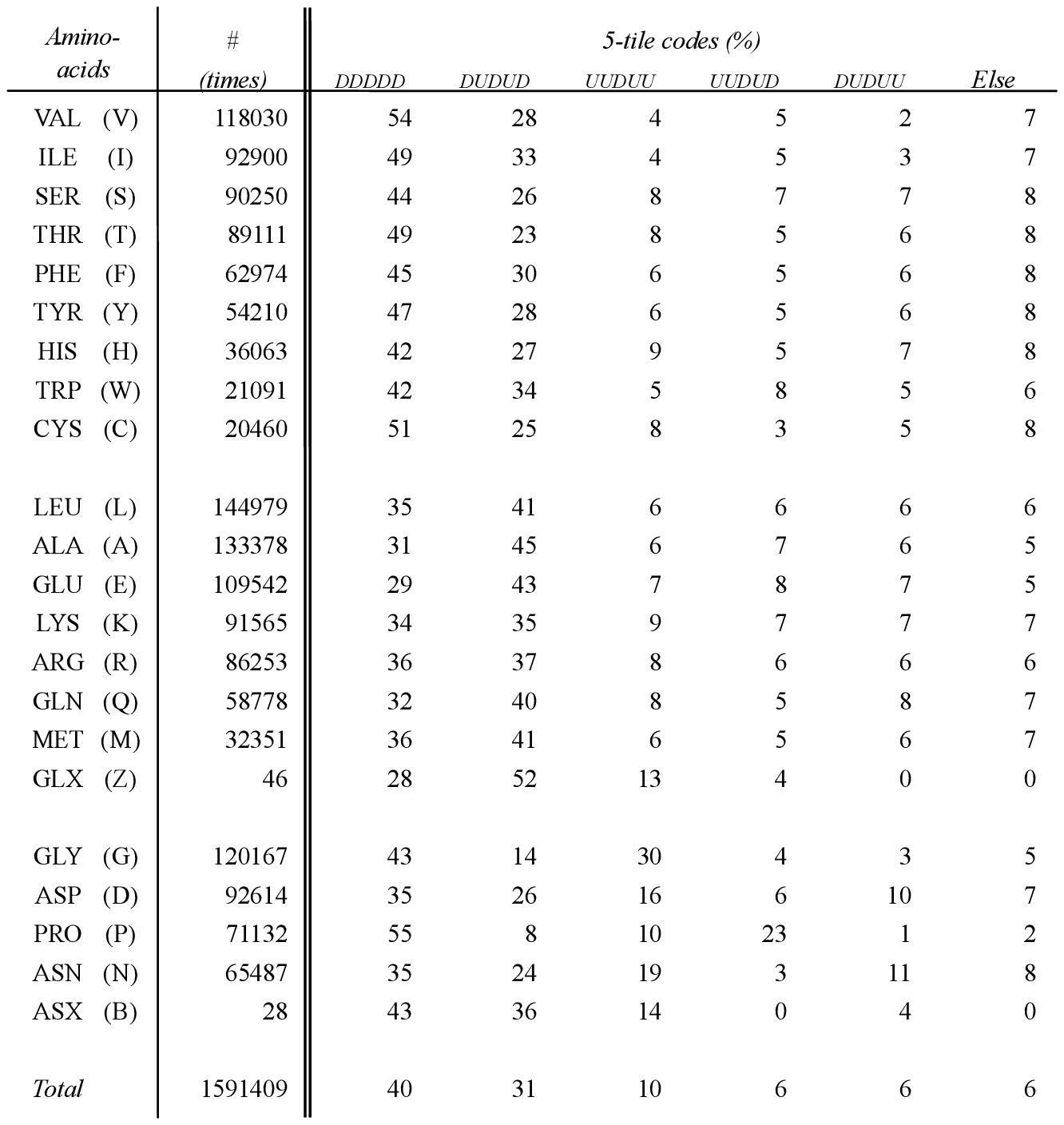}
\caption{Preference of amino-acids for $5$-tile codes.}
\label{tab4}
\end{table*}

\end{document}